\input{aipcheck}
\documentclass[
    ,final            
  ]
  {aipproc}

\layoutstyle{6x9}

\begin{document}

\title{Influence of pions on the hadron-quark phase transition}


\author{O. Louren\c co}{
  address={Departamento de F\'isica, Instituto Tecnol\'ogico de
Aeron\'autica-CTA, 12228-900, S\~ao Jos\'e dos Campos, Brazil}
}

\author{M. Dutra}{
  address={Departamento de F\'isica, Instituto Tecnol\'ogico de
Aeron\'autica-CTA, 12228-900, S\~ao Jos\'e dos Campos, Brazil}
}

\author{T. Frederico}{
  address={Departamento de F\'isica, Instituto Tecnol\'ogico de
Aeron\'autica-CTA, 12228-900, S\~ao Jos\'e dos Campos, Brazil}
}

\author{A. Delfino}{
  address={Instituto de F\'isica, Universidade Federal Fluminense,
Av. Litor\^ anea s/n, 24210-150, Boa Viagem, Niter\'oi RJ, Brazil}
}

\author{M. Malheiro}{
  address={Departamento de F\'isica, Instituto Tecnol\'ogico de
Aeron\'autica-CTA, 12228-900, S\~ao Jos\'e dos Campos, Brazil}
}

\begin{abstract}
In this work we present the features of the hadron-quark phase transition
diagrams in which the pions are included in the system. To construct such
diagrams we use two different models in the description of the hadronic and
quark sectors. At the quark level, we consider two distinct parametrizations
of the Polyakov-Nambu-Jona-Lasinio (PNJL) models. In the hadronic side, we use
a well known relativistic mean-field (RMF) nonlinear Walecka model. We show
that the effect of the pions on the hadron-quark phase diagrams is to move the
critical end point (CEP) of the transitions lines. Such an effect also depends
on the value of the critical temperature ($T_0$) in the pure gauge sector used
to parametrize the PNJL models. Here we treat the phase transitions using two
values for $T_0$, namely, $T_0 = 270$ MeV and $T_0 = 190$ MeV. The last value is
used to reproduce lattice QCD data for the transition temperature at zero
chemical potential.
\end{abstract}

\maketitle



The study of strongly interacting matter is still of great interest in
theoretical and experimental physics \cite{sim}. In particular, this matter is
expected to exhibit a phase transiton at high temperature \cite{nature} or
density \cite{stars} regime. On the theoretical side, Quantum Chromodynamics
(QCD) is the fundamental theory to deal with such problem, but due to its
nonperturbative nature at low energies, a complete description of the
hadron-quark phase transition is not yet possible. As an alternative to the full
QCD description, effective models that share the basic symmetries of QCD can be
used to study this phenomenon (see \cite{eff}, for instance). In this work we
investigate the hadron-quark phase transition in the temperature versus chemical
potential plane, using as the effective quark model the so called
Polyakov-Nambu-Jona-Lasinio (PNJL) model \cite{pnjl,weise,fuku}. For the
hadronic sector, we use a relativistic mean-field (RMF) model including pion
degrees of freedom. Such meson is one of the most abundant particles produced
in the experiments whose goal is the study of the strongly interacting matter
(heavy-ion collisions) \cite{arxiv}. For this reason, it is always important to
analyze its influence also in the context of the hadron-quark phase transition.


It is common to use effective quark models to describe the
hadron-quark phase transition. However, here we use an approach also
applied in Refs. \cite{prd1,prd2}, which considers two distinct
models (a quark and a hadronic one) in the description of the
different phases. For the quark model we use two parameterizations
of the PNJL-SU(2) model whose pressure reads $P_{\mbox{\tiny
PNJL}}=P_q-\mathcal{U}(\Phi,\Phi^*,T)$. The parameterizations are
given by different Polyakov loop potentials
$\mathcal{U}(\Phi,\Phi^*,T)$, which are functions of the Polyakov
loop ($\Phi$) and temperature ($T$). We use the following
parameterizations for $\mathcal{U}(\Phi,\Phi^*,T)$,
\begin{eqnarray}
\frac{\mathcal{U}_{\mbox{\tiny RRW06}}}{T^4} &=& -\frac{b_2(T)}{2}\Phi\Phi^*
+ b_4(T)\mbox{ln}\left[1 - 6\Phi\Phi^* + 4(\Phi^3 + {\Phi^*}^3) -
3(\Phi\Phi^*)^2
\right],\,\,\,\mbox{and} \qquad
\label{rrw06} \\
\frac{\mathcal{U}_{\mbox{\tiny FUKU08}}}{b\,T} &=& -54e^{-a/T}\Phi\Phi^*
+ \mbox{ln}\left[1 - 6\Phi\Phi^* + 4(\Phi^3 + {\Phi^*}^3) - 3(\Phi\Phi^*)^2
\right],
\label{fuku08}
\end{eqnarray}
where $\,b_2(T)=a_0+a_1(T_0/T)+a_2(T_0/T)^2+a_3(T_0/T)^3\,$ and
$\,b_4(T)=b_4(T_0/T)^3$. The parameters $T_0$, $a_i$ ($i=0,1,2,3$), $\,a\,$, and
$\,b\,$, the pressure $\,P_q\,$, as well as the full description of the model,
are found in Ref. \cite{prd1}.

The hadronic model chosen to describe the hadronic phase is the NL3
model \cite{nl3}, added to the free degenerated pion gas. The total
pressure of this system is given by $\,P_{\mbox{\tiny
RMF}}=P_{\mbox{\tiny NL3}}+P_{\mbox{\tiny pions}}\,$ with
\begin{eqnarray}
P_{\mbox{\tiny pions}}=\frac{1}{2\pi^2}\int_0^\infty\frac{k^4dk}{(k^2+m_\pi^2)^{
1/2}}\left\lbrace\frac{1}{e^{[(k^2+m_\pi^2)^{1/2}/T]}-1}\right\rbrace.
\end{eqnarray}
The pion mass is $\,m_\pi=139\,$ MeV. The complete description of the RMF
model, including the $\,P_{\mbox{\tiny NL3}}\,$ definition, is found in Ref.
\cite{prd1}.

\begin{figure}[!htb]
\centering
\includegraphics[height=.35\textheight]{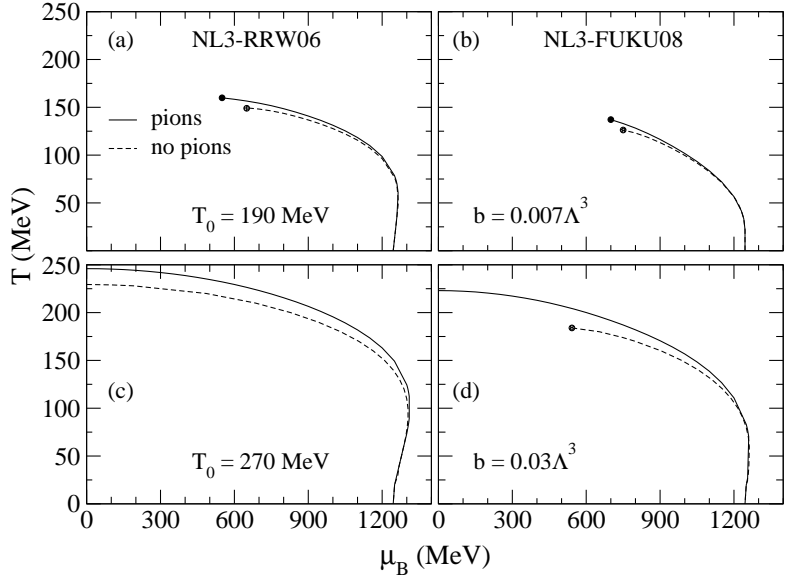}
\caption{Phase diagrams of the matching \mbox{NL3-RRW06} (a-c) and
\mbox{NL3-FUKU08} (b-d) with and without the pions contribution. In panels b
and d $\Lambda=651$ MeV was used. The circles indicate the critical end points.}
\label{figure}
\end{figure}

The matching of the RMF and PNJL models, named as \mbox{RMF-PNJL},
is done by the Gibbs criterium of equal pressures and chemical
potentials at the same temperature. We also use that the baryonic
chemical ($\mu_B$) potential is three times the quark one. The
resulting phase diagram curves are displayed in Fig. \ref{figure}.
\vspace{0.5cm}

In Figs. \ref{figure}a and \ref{figure}b we have used the values of
$\,T_0\,$ and $\,b\,$, respectively for the RRW06 and FUKU08 PNJL
models. Note that, the phase transition lines constructed
exclusively with the PNJL models are consistent with lattice data
\cite{prd1}. In Figs. \ref{figure}c and \ref{figure}d, we have used
the original values of such parameters \cite{weise,fuku}.

Our results show that the main feature of the inclusion of pions in the hadronic
sector of the \mbox{RMF-PNJL} approach, is the change of the critical end point
(CEP) of the transition curves. The position of the CEP is more sensitive to the
inclusion of pions for the \mbox{NL3-FUKU08} hadron-quark phase transition with
the original value of $\,b\,$ parameter, see panel \ref{figure}d. Notice that in
this case, the CEP is completely removed from the transition curve, i. e., the
transition is of first order along all possible values of the baryonic chemical
potential. In the other cases of panels \ref{figure}a, and \ref{figure}b the CEP
position does not change too much. The case depicted in panel \ref{figure}c
reveals that in both calculations (with and without pions) the order of phase
transition remains unchanged. The effect of pions in this case is only the shift
of the transition curve.

As a final remark, we underline that the effect of the pion in the hadron-quark
phase transition is only noticeable at temperatures around $139$ MeV (see Fig.
\ref{figure}). This is a consistent behavior, since the pion degrees of freedom
are expected to become important at $\,T\approx m_\pi$.






\begin{theacknowledgments}
This work was supported by the Brazilian agency
FAPESP (Funda\c c\~ao de Amparo a Pesquisa do Estado de S\~ao Paulo).
\end{theacknowledgments}

\bibliographystyle{aipproc}   

\end{document}